# Computational Design of Metal-Free Porphyrin Dyes for Sustainable Dye-Sensitized Solar Cells Informing Energy Informatics and Decision Support

Md Mahmudul Hasan[1], Chiara Bordin[2]*, Fairuz Islam[3], Tamanna Tasnim[3], Md. Athar Ishtiyaq[3], Md. Tasin Nur Rahim[4], Dhrubo Roy[5]

[1]*Department of Chemistry, Dinajpur Government College, Dinajpur 5200, Bangladesh*

[2]*Department of Computer Science, UiT The Arctic University of Norway, 6050, Langnes, 9037, Tromso, Norway*

[3]*Department of Chemistry, Hajee Mohammad Danesh Science and Technology University, Dinajpur 5200, Bangladesh*

[4]*Department of Aquaculture, Hajee Mohammad Danesh Science and Technology University, Dinajpur 5200, Bangladesh*

[5]*Department of Computer Science and Engineering, Daffodil International University, Dhaka 1216, Bangladesh*

*Corresponding Author: Chiara Bordin
Email: chiara.bordin@uit.no

**Abstract**

This study aims to evaluate the optoelectronic properties of metal free porphyrin-based D-π-A dyes via in-silico performance investigation notifying energy informatics and decision support. To develop novel organic dyes, three acceptor/anchoring groups and five donating groups were introduced to strategic positions of the base porphyrin structure, resulting in a total of fifteen dyes. The singlet ground state geometries of the dyes were optimized utilizing density functional theory (DFT) with B3LYP and the excited state optical properties were explored through time-dependent DFT (TD-DFT) using the PCM model with tetrahydrofuran (THF) as solvent. Both DFT and TD-DFT calculations were carried out using the 6-311G(d,p) basis set. The HOMO energy levels of almost all the modified dyes are lower than the redox potential of $I^-/I_3^-$ and LUMO energy levels are higher than the conduction band of $TiO_2$. The absorption maxima values ranged from 690.64 to 975.55 nm. The dye N1 using triphenylamine group as donor and p-ethynylbenzoic acid group as acceptor, showed optimum optoelectronic properties ($\Delta G_{reg}$=-9.73 eV, $\Delta G_{inj}$=7.18 eV, $V_{OC}$=1.47 V and $J_{SC}$=15.03 mA/cm$^2$) with highest PCE 14.37%, making it the best studied dye. This newly modified organic dye with enhanced PCE is remarkably effective for the dye-sensitized solar cells (DSSC) industry. Beyond materials discovery, this study highlights the role of high-performance computing in enabling predictive screening of dye candidates and generating performance indicators (HOMO–LUMO gaps, absorption spectra, charge transfer free energies, photovoltaic metrics). These outputs can serve as key parameters for energy informatics and system modelling.

**Introduction**

Dye-sensitized solar cells (DSSCs) are a type of solar cell that convert sunlight into electricity using dye molecules on a semiconductor surface. These devices are valued for their low-cost manufacturing process, simple fabrication procedure and their reliability on non-toxic materials (Malhotra et al., 2024). DSSCs can be made even more eco-friendly and affordable, by using metal-free organic dyes instead of rare or toxic metal complexes. These organic sensitizers capture a broad spectrum of sunlight (from ultraviolet to near infrared) and have yielded impressive efficiencies which can be validated by studies in which all-organic DSSC systems had reached about 12.45% power conversion efficiency (Khadiri et al., 2025). In recent times, porphyrin derivatives have emerged as especially promising metal-free sensitizers for DSSCs. Porphyrin derivatives provide strong thermal and electronic stability along with excellent light absorption properties, making them well suited for solar applications (Birel et al., 2017). In practice, porphyrin dyes have achieved efficiencies on par with the best metal-based dyes: for example, a porphyrin sensitizer was shown to give 11.5% power conversion efficiency (Abdel-Latif et al., 2023). In fact, an engineered free-base porphyrin (SM315) achieved a record 13% efficiency in a DSSC, surpassing many earlier organic dyes (Mathew et al., 2014). These results highlight that metal-free porphyrins can replace low-efficiency dyes and significantly boost DSSC performance.

A key reason for this high performance is the ability to extend π-conjugation and tune donor-acceptor interactions in the porphyrin core. The conjugated system lengthens and intramolecular charge transfer (ICT) is enhanced by attaching strong electron-donating and electron-accepting groups to the porphyrin. For example, adding an N, N-dimethylamino donor group to a porphyrin increased ICT character and greatly improved the dye's light-harvesting efficiency (Kang et al., 2016). Similarly, the SM315 porphyrin's donor-π-acceptor design maximizes electron delocalization and red-shifts its absorption, which greatly improve light harvesting efficiency (Mathew et al., 2014). Such modifications raise the oscillator strength of the main electronic transitions and extend absorption toward the red/near-IR region, improving how much sunlight the dye can absorb.

Recently, several studies have used DFT as an independent tool to study the material properties of solar cells. Giorgi et al. (2014) used DFT analysis to investigate the role played by the MA cation in determining the band structure of MAPbI3. Ma et al., (2014) on the other hand, developed a set of algorithms based on DFT and time-dependent DFT calculations to accurately predict the PCE of dye-sensitized solar cells. The DFT and TD-DFT simulations were carried out using high-performance computing (HPC) resources, which enabled efficient evaluation of multiple candidate dyes. In this study, we examined the influence of modifying five donor groups and three anchoring groups on the geometries and optoelectronic properties (HOMO, LUMO, $\Delta G_{reg}$, $\Delta G_{inj}$, $V_{OC}$, PCE and etc.) of fifteen organic dyes with a D-π-A structure using the methods DFT and TD-DFT. The selection of these donor moieties is substantiated by their remarkable performance in similar DSSCs (Li et al., 2014; Le et al., 2017; Kumar et al., 2018; Yuan et al., 2021; Bessho et al., 2010). We systematically examined the effect of substituting the studied donor units with three distinct anchoring groups (p-ethynylbenzoic acid, p-tolyl phosphonic acid and p-thiophosphonic acid group) on the optoelectronic attributes. The goal was to improve the optoelectronic properties of these modified dyes while maintaining a balanced synergy with the semiconductor and redox electrolyte $I^-/I_3^-$.

Overall, the DFT/TD-DFT screening of fifteen D-π-A porphyrin dyes shows that, deliberate tuning of donor strength and choice of anchoring group can significantly modulate frontier orbital energies, strengthen intramolecular charge transfer, red-shift of absorption spectrum and improve the thermodynamics for electron injection along with dye regeneration. Framing this workflow as a computational pipeline (design → optimization → property prediction → performance metrics) highlights its scalability: such in-silico approaches can be extended to larger dye libraries and integrated with data-driven methods such as machine learning. This positions the present work within the broader field of computational sustainability, where computational methods are used to accelerate discovery while reducing costly trial-and-error experimentation. This approach bridges computational chemistry with energy informatics, showing how material-level modelling can feed into system-level decision making. The computationally identified donor-acceptor/anchor combination positions the work within the broader domain of computational sustainability, where computational resources are leveraged to accelerate innovation in sustainable energy technologies. While most computational studies stop at reporting molecular properties, this work explicitly frames these results as inputs to system-level models, providing a bridge between materials discovery and energy transition planning. Follow-up work should prioritize synthesis of the top candidates, measuring their optoelectrical properties and testing complete devices under standard conditions to confirm the current work predictions, while evaluating photochemical and thermal stability. Finally, combining these optimized molecules with interface engineering, semiconductor morphology control and electrolyte optimization will be key to translating the molecular advantages into robust and environmentally friendly DSSCs.

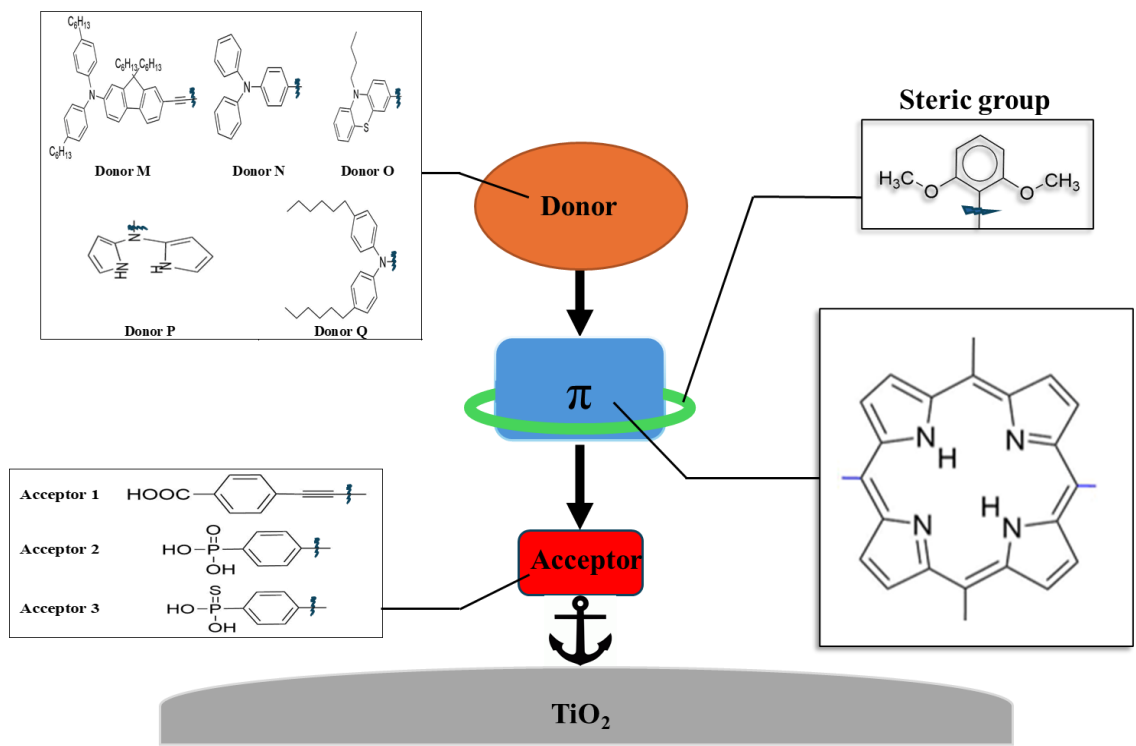

**Fig. 1** Molecular structures of representative porphyrin-based donor, π and acceptor materials for effective molecular engineering strategies

## Methods

### Dye Design and Evaluation

All the molecular structures were initially constructed using GaussView 6.0 software. The calculations were performed using the Gaussian16 software package (Frisch et al., 2016). The simulations were executed on a high-performance computing (HPC-class) workstation (Dell Precision 7875 Tower with an AMD Ryzen Threadripper PRO 7965WX processor, 24 cores/48 threads, 128 GB ECC DDR5 RAM, and NVMe SSD storage). This HPC-class environment provided the computational capacity to handle demanding DFT and TD-DFT workflows and enabled efficient evaluation dye candidates. While the present study focused on a limited set of dyes, the availability of such resources makes it possible to scale in-silico screening pipelines to much larger molecular libraries, demonstrating the importance of HPC resources in advancing computational sustainability for large-scale in-silico screening of photovoltaic materials discovery.

In this study, a library of fifteen dyes was constructed and stratified into five series (M, N, O, P, Q) according to the nature of the donor moiety. Within each series, three distinct derivatives were crafted by varying the acceptor/anchoring groups into the core porphyrin framework, denoted by numerical suffixes 1, 2 and 3, yielding the sets M1-M3, N1-N2, O1-O3, P1-P3, and Q1-Q3 (Fig. 1). Ground state geometry optimizations were performed using a density functional theory (DFT) approach at B3LYP/ 6-311G (d, p) level of theory. The basis set 6-311G (d, p) performed well enough to capture the frontier molecular orbital energies. All optimized structures were confirmed to be true minima by the absence of imaginary frequencies in subsequent vibrational frequency calculations. To investigate the electronic absorption properties, time dependent-density functional theory (TD-DFT) calculations were carried out at the same level of theory, B3LYP/6-311G (d, p). Solvent effect was included in the calculation of excited states which is required in depicting the real environment of the system using polarizable continuum method (PCM) with tetrahydrofuran (THF) as the solvent (Santhanamoorthi et al., 2013; Tomasi et al., 2005).

Open-circuit photo-voltage ($V_{OC}$) and short-circuit current density ($J_{SC}$) are two essential metrics used to evaluate the power conversion efficiency (PCE) of DSSCs (Chenab et al., 2019). A foundation for designing high conversion-efficiency DSSCs is provided by the PCE and can be determined by theoretical methods using equation 1 (Bourouina & Rekhis, 2017)

$$PCE = FF \frac{J_{SC} \times V_{OC}}{P_{in}} \times 100\% \qquad (1)$$

Where FF is the fill factor and P$_{in}$ is the incident solar power on the cell. The incident power from the incoming light source is inversely corelated with the photovoltaic material's PCE. It is known that a molecule's band gaps and charge transfer rates are significantly influenced by its J$_{SC}$ value (Hussain et al., 2022). The value of J$_{SC}$ in DSSCs is determined as the equation 2 (Sun et al., 2016):

$$J_{SC} = q \int LHE(\lambda)\, \phi_{inj} \eta_{coll} d\lambda \qquad (2)$$

Where q indicates the charge of the electron, LHE is light-harvesting efficiency, $\eta_{coll}$ stands for the charge collection efficiency that is equivalent to a constant for the same DSSC architecture using only distinct sensitizers and $\phi_{inj}$ is the electron injection efficiency.

The light harvesting efficiency LHE must be as larger as feasible in order to get the highest possible J$_{SC}$. LHE is calculated from equation 3 (Zhang et al., 2010) (Zhang et al., 2018):

$$LHE(\lambda_{max}) = 1 - 10^{-f} \qquad (3)$$

Where $f$ indicates the oscillator strength corresponding to the wave length $\lambda_{max}$ of the dye.

The free energy of electron injection ($\Delta G_{inj}$) used to estimate the electron injection efficiency ($\phi_{inj}$), which can be calculated as the equation 4 (Shi et al., 2019):

$$\Delta G_{inj} = (E_{dye*} - E_{CB}) \qquad (4)$$

Where, $E_{dye*}$ is the oxidation potential energy of the dye in the excited state and $E_{CB}$ is the conduction band (CB) of TiO$_2$.

The oxidation potential energy of the excited state can be calculated using equation 5:

$$E_{dye*} = E_{dye} - E \qquad (5)$$

Where $E_{dye}$ is the oxidation potential at the ground state and the electronic vertical transition energy that corresponds to $\lambda_{max}$ is represented by $E$. Based on Koopman's theorem, $E_{dye}$ can be estimated by taking negative of the dye E$_{HOMO}$ (Pearson, 1988).

The driving force regeneration $\Delta G_{reg}$ for the dye is calculated by the following equation 6 (J. Zhang et al., 2017):

$$\Delta G_{reg} = (E_{dye} - E_{I^-/I_3^-}) \qquad (6)$$

Where, $E_{I^-/I_3^-}$ is the redox potential energy of the redox electrolyte.

Apart from the short-circuit density, the open-circuit voltage $V_{OC}$ is an additional crucial factor that determines the conversion efficiency of a DSSCs. The $V_{OC}$ can be estimated by the equation 7 (Barati-Darband et al., 2018):

$$V_{OC} = E_{LUMO} - E_{CB}^{TiO_2} \qquad (7)$$

Where, $E_{CB}^{TiO_2}$ denotes the energy of TiO$_2$ semiconductor conduction band and $E_{LUMO}$ is the LUMO energy level of the dye.

Some parameters of the several organic dyes under study were calculated in order to compare their reactivity and stability, such as ionization potential (IP), electron affinity (EA), chemical hardness(η), electrophilicity ($\omega$), electron-donating power ($\omega^-$) and electron-accepting power ($\omega^+$). The ionization potential (IP) and electron affinity (EA) are approximated as the negative of the HOMO and LUMO energies respectively (Pearson, 1988[b]).

$$IP = -E_{HOMO} \qquad (8)$$

$$EA = -E_{LUMO} \qquad (9)$$

The equation (8) and (9) show an alternative method for estimating the IP and EA (Y. Li et al., 2016).

The chemical hardness (η), electrophilicity ($\omega$), electron-donating power ($\omega^-$) and electron accepting power ($\omega^+$) of dyes were computed by using the following equations (Martínez et al., 2012) (Gázquez et al., 2007):

$$\eta = (IP - EA)/2 \qquad (10)$$

$$\omega = \frac{(IP+EA)^2}{8(IP-EA)} \qquad (11)$$

$$\omega^- = \frac{(3IP+EA)^2}{16(IP-EA)} \qquad (12)$$

$$\omega^+ = \frac{(IP+3EA)^2}{16(IP-EA)} \qquad (13)$$

**Computational Workflow and Conceptual System-Level Framework**

The study utilizes a robust in-silico modelling pipeline for screening dye candidates, centered on Density Functional Theory (DFT) and Time-Dependent DFT (TD-DFT) simulations. The workflow includes geometry optimization, frontier orbital analysis (HOMO–LUMO), and optical property predictions (absorption spectra, photovoltaic parameters). The sequence of tasks (design → optimization → property prediction → performance metrics) can be conceptualized as a computational workflow. Such workflows align with the logic of informatics pipelines in other domains (e.g., bioinformatics), providing structured, repeatable analysis to support decision-making.

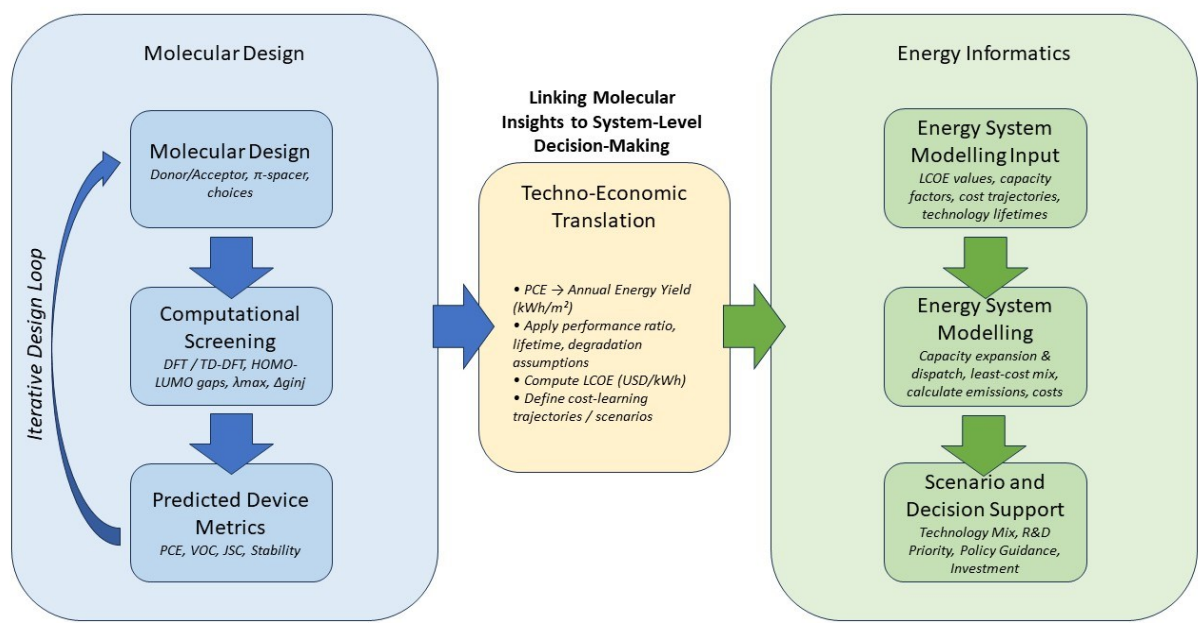

**Fig. 2** Conceptual integration pipeline linking molecular design to energy informatics and decision support. While the techno-economic translation step is illustrated with idealized assumptions, it demonstrates how predicted PCE values could be converted into LCOE estimates and cost-learning trajectories for future integration into energy system models.

The workflow can in principle be scaled up to screen large libraries of dyes, generating datasets of predicted performance. These datasets may serve as training material for machine learning models, enabling predictive analytics and automated discovery in photovoltaic research. This Computational screening reduces costly trial-and-error in the lab, thereby improving efficiency and sustainability in materials development. By narrowing the candidate set before synthesis, the computational approach exemplifies computational sustainability and creates a natural link to prescriptive analytics, guiding which dyes to prioritize. Ultimately, this computational methodology ensures efficiency and sustainability by replacing resource-intensive wet-lab trial-and-error with prescriptive analytics, guiding the prioritization of promising dye structures.

The Fig. 2 illustrates the conceptual computational workflow proposed in this study, highlighting how molecular-level simulations can inform system-level decision-making. In the figure we propose a computational pipeline that can generate early-stage performance data suitable for use in techno-economic and energy system models. On the left, the molecular design domain consists of an iterative loop in which novel porphyrin dyes are designed, screened via DFT/TD-DFT simulations, and evaluated through quantitative performance indicators (HOMO-LUMO gaps, $\lambda_{max}$, $\Delta G_{inj}$, $V_{oc}$, $J_{sc}$, PCE). The curved arrow highlights that these indicators are used to guide subsequent design iterations, creating a feedback loop that continuously improves candidate selection. The central two-colored arrow emphasizes the potential for bridging molecular insights (blue) with energy informatics and decision support (green): the indicators can serve as input parameters for energy informatics applications, such as energy system modelling, technology foresight, and decision-support frameworks, linking molecular-level design to system-level strategy for sustainable energy transitions. On the right, the energy informatics domain contextualizes the computational results by translating them into energy system modelling inputs, enabling techno-economic and scenario analysis, and ultimately supporting R&D prioritization, investment focus, and policy guidance. Together, the figure highlights that this approach is not merely molecular modelling but a pipeline that links materials discovery to strategic decision-making for sustainable energy transitions.

**Results and Discussions**

**Structure of Dyes**

The designed porphyrin dyes as shown in Figure 1 can be characterized by a donor-π-acceptor (D-π-A) configuration, where the central porphyrin core serves as the π-conjugated bridge. Among the four meso positions of the porphyrin ring, two sterically bulky groups are attached at two opposite meso positions to reduce aggregation and influence the dye's orientation on the $TiO_2$ surface (Vaz and Pérez-Lorenzo 2023). Additionally, another two positions are functionalized with one electron donating group and one electron withdrawing group to facilitate efficient charge transfer. Total five of effective donor groups were used for modification. The acceptor groups employed include benzoic acid, pyridyl phosphorothioic acid and pyridyl phosphonic acid. These groups not only serve as electron withdrawing entities but also act as anchoring group due to the presence of oxygen atoms which are capable of bind with the $TiO_2$ surface. A total of fifteen newly porphyrin-based metal free dyes were modified using the combination of these electron donating groups and electron withdrawing groups. These dyes produce electrons by the molecule's donor's higher ability to donate electron and acceptor's ability to withdraw electron which also improves the intramolecular charge transfer capabilities of the dyes (Wang et al., 2018). These dyes work as superior sensitizers in DSSCs because the meso position of porphyrin system is highly active in functionalizing different anchoring groups (Krishna et al., 2019).

**Frontier Molecular Orbitals**

The frontier molecular orbitals (FMOs) were studied to understand the molecule's electronic structure and kinetic stability by the interpretation of widely used quantum chemical descriptors like, the highest occupied Molecular orbital (HOMO) and lowest unoccupied molecular orbital (LUMO). It also gives insight into the reactivity, regio selectivity of various molecules (Woodward & Hoffmann, 1969). A HOMO-LUMO gap reflects greater electronic delocalization and facilitates easier charge transfer within the system (Li et al., 2016). This also play a decisive role in determining the light-harvesting of dye sensitizers in dye-sensitized solar cells (DSSCs). The HOMO energy level primarily influences the dye's regeneration process by redox couple, whereas the LUMO level governs the injection of photoexcited electrons into the conduction band (CB) of the semiconductor, typically $TiO_2$.

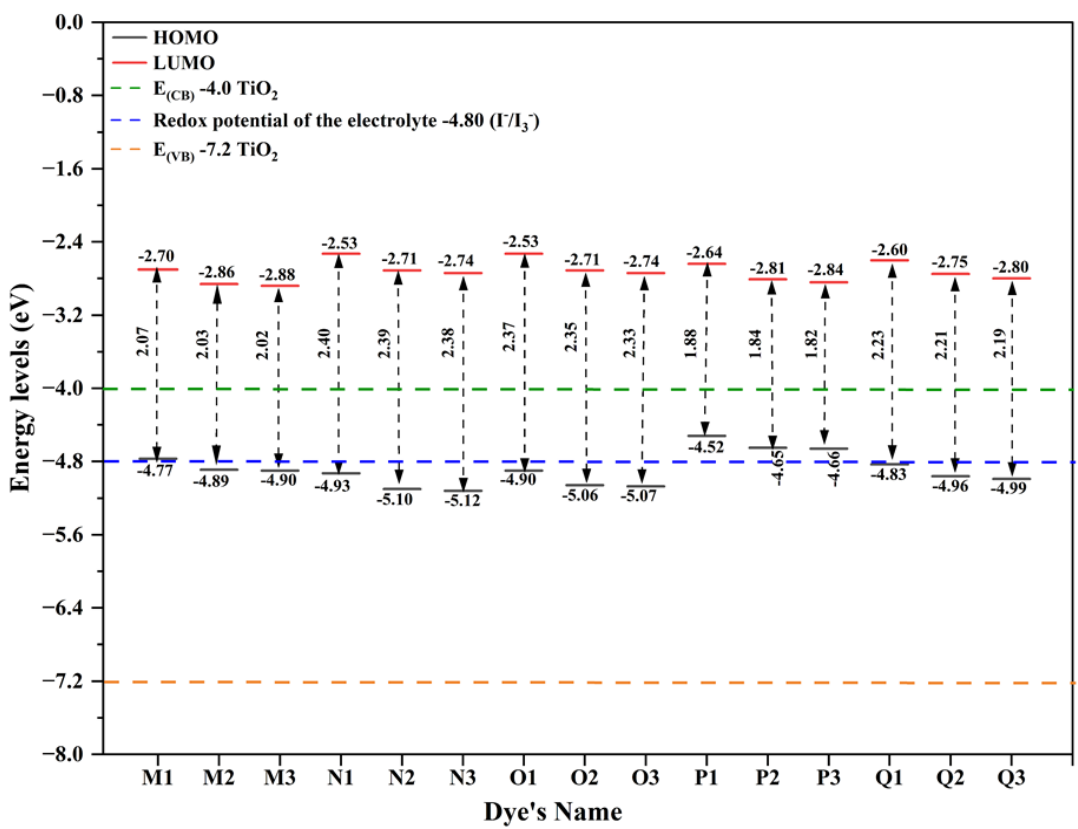

**Fig. 3** Highest occupied molecular orbital (HOMO) and lowest unoccupied molecular orbital (LUMO) energy levels of porphyrin-based dyes

For efficient electron injection, the LUMO energy of the dye must be more than the conduction band of $TiO_2$ (-4.0 eV), facilitating spontaneous electron transfer upon photoexcitation (Divya and Suresh 2021). The HOMO energy should be positioned lower than the redox potential of the electrolyte ($I^-/I_3^-$) to aid in the quick regeneration of dye (Arunkumar and Anbarasan 2023). In the present study, the frontier molecular orbitals of M, N, O, P and Q series dyes are illustrated in Figure 3. Energy band gap is known to be a good indicator to determine the efficiency of a dye (Adnan et al., 2023). The calculated HOMO energies of the designed dyes range from -4.52 to -5.12 eV, while the LUMO energies vary between -2.53 to -2.88 eV. All structures display LUMO levels significantly higher than the CB of $TiO_2$, suggesting that electron injection from the dye into the semiconductor is thermodynamically favorable. Furthermore, the HOMO level of most of the dyes are sufficiently lower than the redox potential of the $I^-/I_3^-$ couple, indicating that the regeneration of the dyes is also feasible. The HOMO-LUMO gaps of efficient candidates vary between 2.02 eV (M3) and 2.40 eV (N1). According to the literature analysis, a photovoltaic material with a smaller band gap will exhibit a higher PCE. From Figure 3 and 4 it can be inferred that the strong electron withdrawing acceptor, pyridyl phosphorothioic acid, is responsible for the lowest band gap among the studied dyes. Furthermore, the pyridyl phosphoric acid moiety also exhibits a smaller band gap compared to the commonly used benzoic acid acceptor. The maximum EA and minimum IP can be observed in the Table 1. Among the efficient candidates the minimum IP can be observed for Q1, which indicates that it is easier to release electrons to create holes for Q1 than the other efficient dyes.

**Table 1** Ground states $E_{HOMO}$, $E_{LUMO}$, energy gap (Eg), ionization potential (IP) and electron affinity

| Dyes | $E_{HOMO}$ (eV) | $E_{LUMO}$ (eV) | Energy gap ($E_g$) (eV) | IP (eV) | EA (eV) |
|---|---|---|---|---|---|
| M1 | -4.77 | -2.70 | 2.07 | 4.77 | 2.70 |
| M2 | -4.89 | -2.86 | 2.03 | 4.89 | 2.86 |
| M3 | -4.90 | -2.88 | 2.02 | 4.90 | 2.88 |
| N1 | -4.93 | -2.53 | 2.40 | 4.93 | 2.53 |
| N2 | -5.10 | -2.71 | 2.39 | 5.10 | 2.71 |
| N3 | -5.12 | -2.74 | 2.38 | 5.12 | 2.74 |
| O1 | -4.90 | -2.53 | 2.37 | 4.90 | 2.53 |
| O2 | -5.06 | -2.71 | 2.35 | 5.06 | 2.71 |
| O3 | -5.07 | -2.74 | 2.33 | 5.07 | 2.74 |
| P1 | -4.52 | -2.64 | 1.88 | 4.52 | 2.64 |
| P2 | -4.65 | -2.81 | 1.84 | 4.65 | 2.81 |
| P3 | -4.66 | -2.84 | 1.82 | 4.66 | 2.84 |
| Q1 | -4.83 | -2.60 | 2.23 | 4.83 | 2.60 |
| Q2 | -4.96 | -2.75 | 2.21 | 4.96 | 2.75 |
| Q3 | -4.99 | -2.80 | 2.19 | 4.99 | 2.80 |

The distribution of frontier molecular orbitals is an important factor that affects the intramolecular charge transfer of dyes and also affects the performance of DSSCs. Figure 3 shows the distributions of the FMOs of the dyes. The LUMOs are mainly concentrated in the π linker and anchor acceptor. The HOMOs are mostly concentrated in the electron donor and less concentrated in the porphyrin ring. This makes dyes to achieve better electron-hole separation and promote intra molecular charge transfer. The electron density distribution confirms efficient charge transfer from the donor to the acceptor, followed by electron injection into the conduction band (CB) of $TiO_2$.

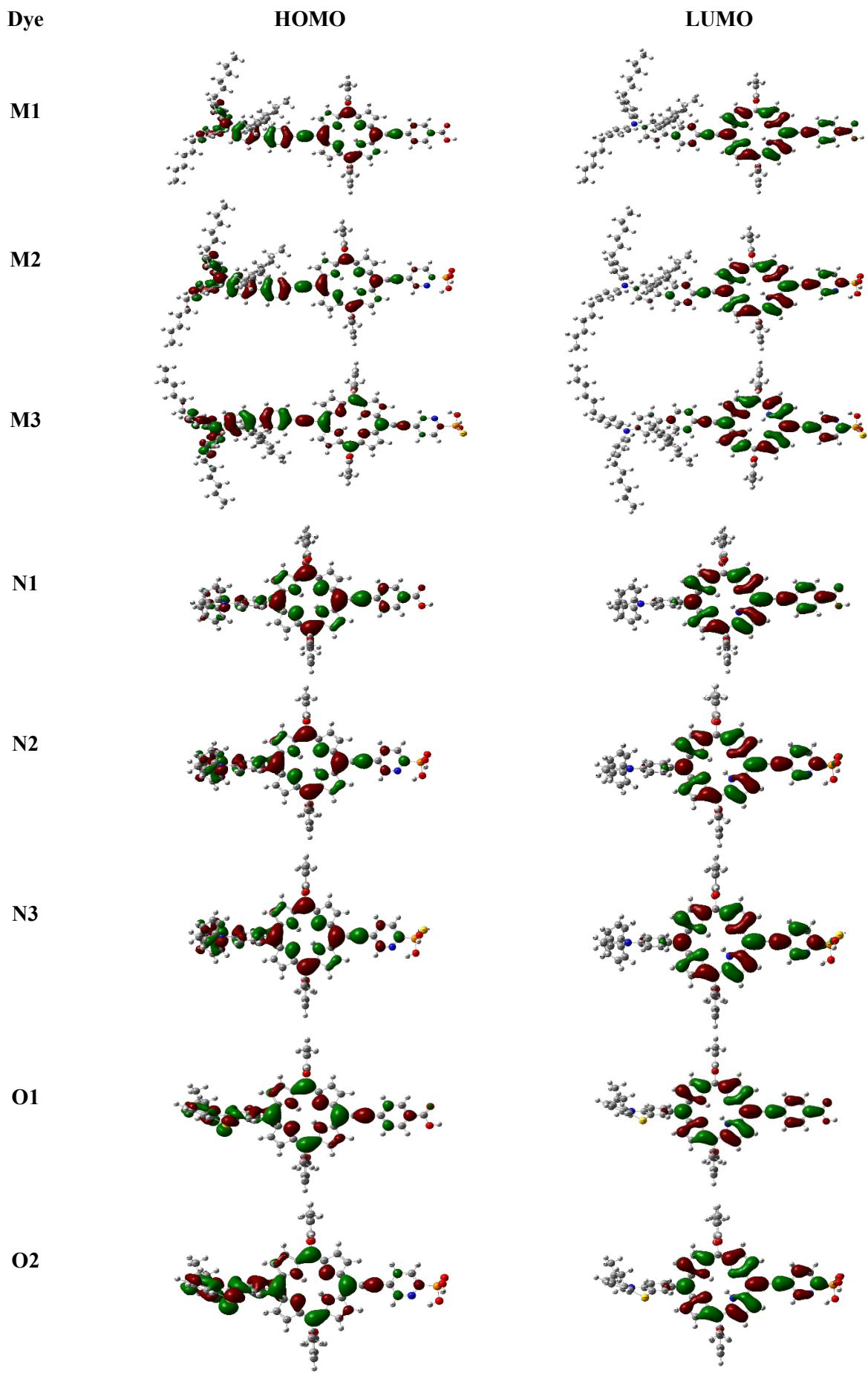

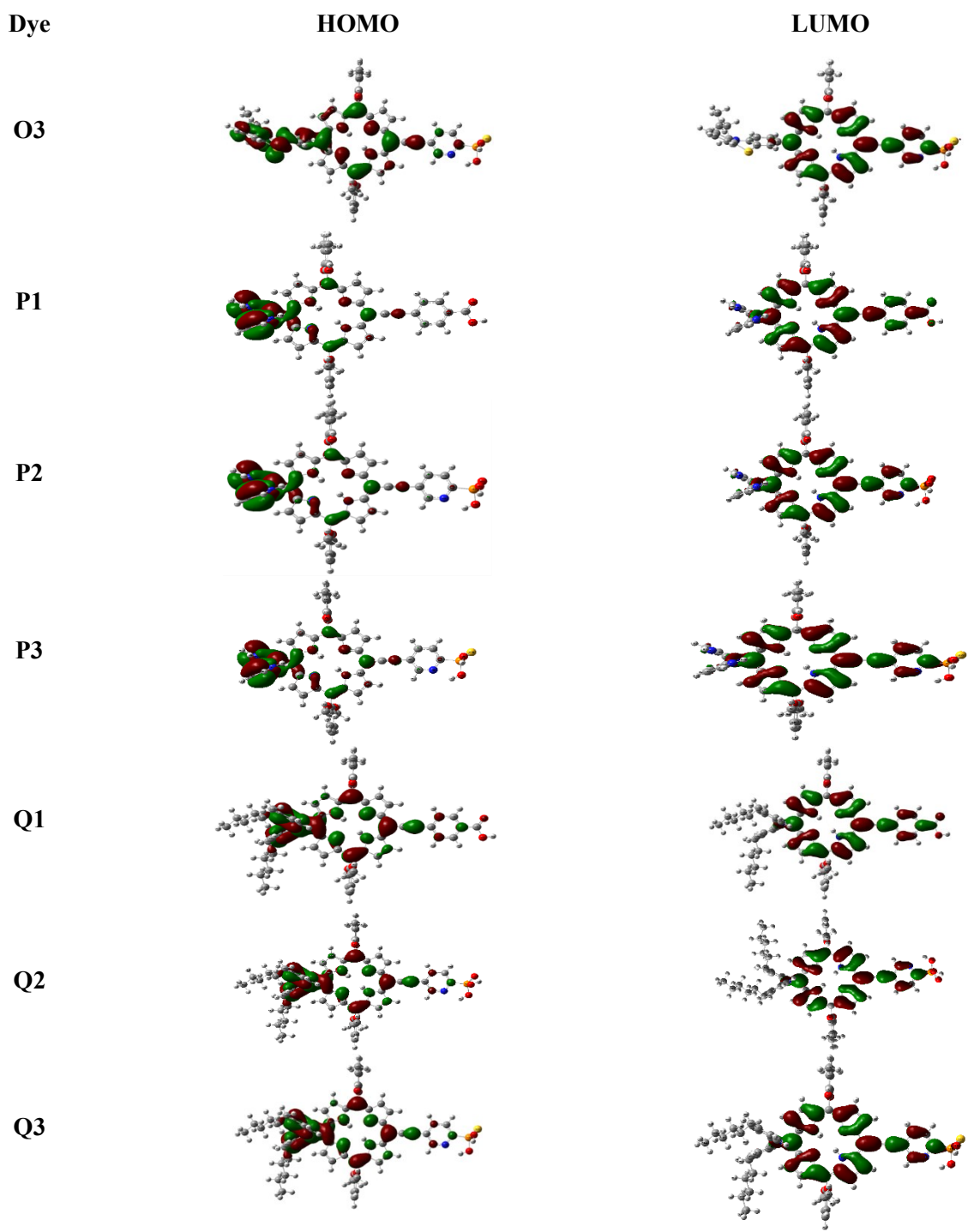

**Fig. 4** Frontier molecular orbitals of dyes obtained with B3LYP/6-311G(d,p)

**UV-Vis Absorption Spectra**

The simulated absorption spectrum provides insight into both the absorption intensity, oscillator strength and wavelength range of the dyes toward incident photons (Figure 5). A broadened absorption profile enhances the light-harvesting capability of the dyes (Abdullah et al., 2022). Molecules that exhibit pronounced absorption bands together with high oscillator strengths are considered suitable candidate for photovoltaic applications (Setsoafia et al., 2024).

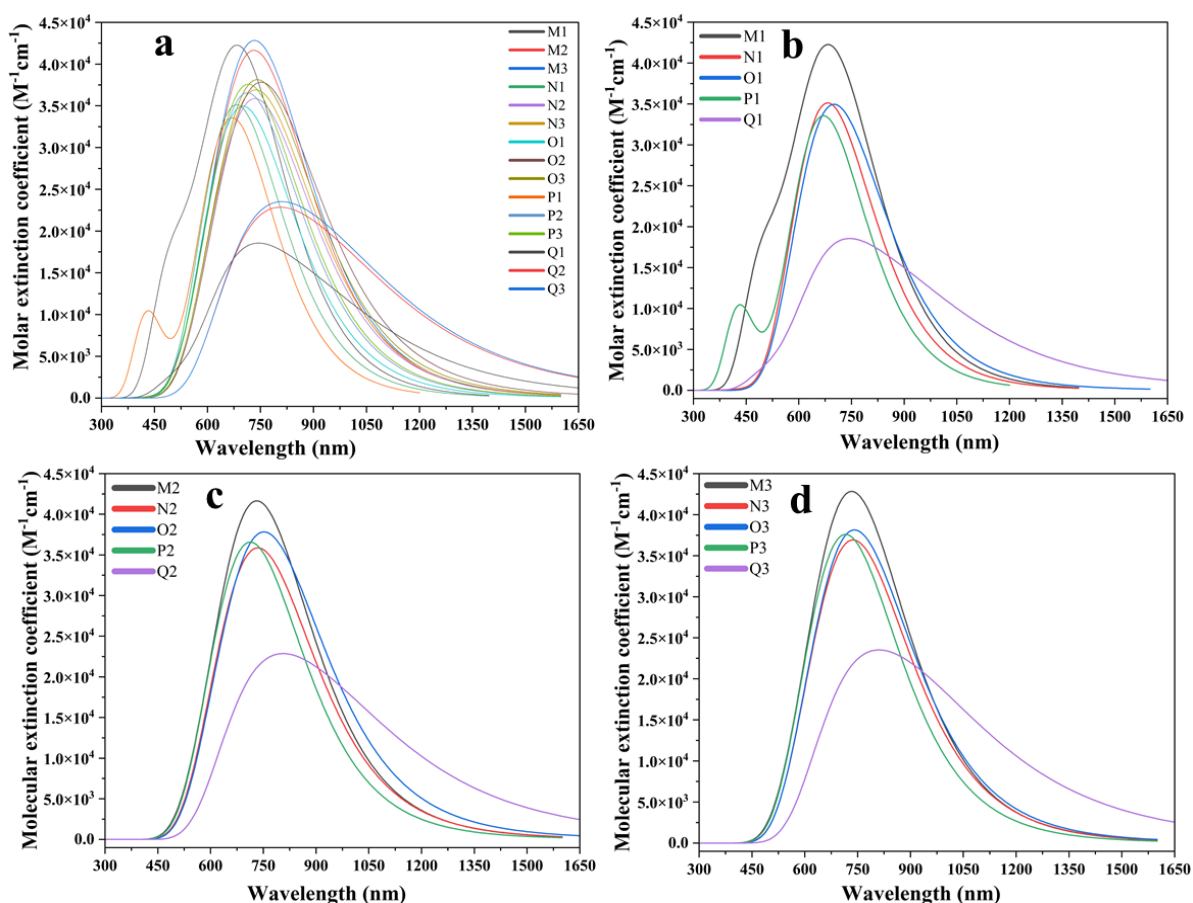

**Fig. 5** Ultraviolet-Visible (UV-Vis) absorption spectra of porphyrin-based dyes; a) UV-Vis Spectra of all dye, b) UV-Vis Spectra of M1, N1, O1, P1 and Q1, c) UV-Vis Spectra of M2, N2, O2, P2 and Q2, d) UV-Vis Spectra of M3, N3, O3, P3 and Q3.

The absorption wavelengths, vertical excitation energy, oscillator strength and the orbitals involved in the transition of porphyrin dyes are presented in Table 2. Expanding absorption into the visible or near infrared ray (NIR) region enables the dye to capture more solar photons, which boosts photocurrent generation and efficiency of the DSSCs. Previous research observed a red shift in $\lambda_{max}$ when extending π-conjugation in D-π-A system, illustrating that a longer conjugated path enhances visible light absorption which is a key design strategy in DSSCs (Kargeti et al., 2023). In Table 2 it is clearly shown that the $\lambda_{max}$ values ranging from 690.64 nm to 975.55 nm, which indicates the intense absorption bands in the visible to NIR region where Q3 shows the most red-shifted absorption at 975.55 nm. Q1, Q2 and Q3 shows broader peaks but lower molecular extinction coefficient. M2, M3, N1, N2 and N3 shows moderate values of absorption maxima (755.45 nm, 756.63 nm, 708.70 nm, 767.47 nm and 767.49 nm, respectively). M3, N2 and N3 have the highest, H →L transitions and highest to moderate oscillator strength (*f*)*,* H →L (100%) with *f* = 0.9076 for M3, H →L (100%) with *f* =0.7277 for N2, H →L (100%) with *f* =0.7277 for N3. These results suggest the effectiveness of M series dyes in terms of molar extinction coefficient while Q series dyes showed the most red-shifting. However, N series dyes shower a moderate molecular extinction coefficient with λ$_{max}$ values lying in between 708.70 to 769.49 nm wavelength, making them an optimum series of dyes in terms of photon absorption.

**Table 2** Absorption wavelengths (λ$_{max}$), vertical excitation energy (E), oscillator strength (*f*) and the orbitals involved in the transition of porphyrin dyes

| Dyes | λ$_{max}$ (nm) | E (eV) | *f* | Transition H = HOMO, L = LUMO (%) |
|---|---|---|---|---|
| **M1** | 709.64 | 1.74 | 0.8096 | H-1→L (99.99%)<br>H→L (94.83%); H→L+1 (89.20%) |
| **M2** | 755.45 | 1.64 | 0.8802 | H→L (99.99%); H→L+2 (99.99%)<br>H→L+1 (88.15%); H-3→L (9.04%) |
| **M3** | 756.63 | 1.63 | 0.9076 | H→L (100%); H→L+2 (99.99%)<br>H →L+1 (88.35%); H-3→L (8.92%) |
| **N1** | 708.70 | 1.75 | 0.605 | H→L (95.61%); H→L+1 (88.37%)<br>H -2→L (59.50%); H-1→L (40.49%) |
| **N2** | 767.47 | 1.61 | 0.7036 | H→L (100%); H→L+2 (99.99%)<br>H→L+1 (90.77%); H -2→L (9.22%) |
| **N3** | 769.49 | 1.61 | 0.7277 | H→L (100%); H→L+2 (100%)<br>H→L+1 (91.78%); H -2→L (8.21%) |
| **O1** | 743.97 | 1.66 | 0.5263 | H→L (87.99%); H -2→L (86.06%)<br>H-1→L (84.59%); H→L (12.31%) |
| **O2** | 793.88 | 1.56 | 0.6604 | H→L+1 (87.21%); H→L (85.72%)<br>H-1→L (71.71%); H -1→L (14.27%) |
| **O3** | 777.76 | 1.59 | 0.6721 | H→L+1 (89.64%); H→L (86.74%)<br>H-1→L (84.34%); H-1→L (12.43%) |
| **P1** | 690.64 | 1.79 | 0.5019 | H→L (96.10%); H→L+2 (91.79%)<br>H→L+1 (89.69%); H-1→L (10.30%) |
| **P2** | 751.16 | 1.65 | 0.6318 | H→L (97.59%); H→L+2 (97.51%)<br>H→L+1 (92.39%); H-1→L (5.43%) |
| **P3** | 753.47 | 1.64 | 0.656 | H→L (97.52%); H→L+2 (95.24%)<br>H→L+1 (94.15%); H-1→L (5.84%) |
| **Q1** | 932.67 | 1.33 | 0.2389 | H→L (97.86%); H→L+2 (95.66%)<br>H→L+1 (91.91%); H-2→L (8.08%) |
| **Q2** | 974.94 | 1.27 | 0.3183 | H→L (99.99%); H→L+1 (97.64%)<br>H→L+2 (94.29%); H-2→L (5.70%) |
| **Q3** | 975.55 | 1.27 | 0.3283 | H→L (99.99%); H→L+1 (97.86%)<br>H→L+2 (94.22%); H-2→L (5.77%) |

**Chemical reactivity parameters**

The most stable dyes are those with the highest chemical hardness ($\eta$), whereas low $\eta$ (soft) dyes are more easily polarized and participate in charge transfer (Kaya and Putz 2022). In Table 3, the values of $\eta$ ranges from 0.91 to 0.94 eV for P series dyes which can be considered the softest among all the dye groups. However, N and O series dyes showed the most hardness with the values ranging between 1.17 to 1.20 eV. Conceptual DFT studies of DSSC dyes consistently show that low $\eta$ correlates with enhanced intramolecular charge transfer and higher efficiency (Pakravesh et al., 2021). Some studies suggest that, dyes engineered for minimal hardness tend to exhibit the highest conversion efficiency (Delgado-Montiel et al., 2020). In the current work, P and M series were found to be the softest dyes, implying facile electron donation to $TiO_2$. On the contrary, the N and O series are more resistant to oxidation and potentially more robust.

**Table 3** Chemical reactivity parameters of porphyrin-based dyes (in eV) obtained by density functional theory (DFT) conceptual at B3LYP/6-311G(d,p) level of theory

| Dyes | $\eta$ | $\omega$ | $\omega^-$ | $\omega^+$ |
|---|---|---|---|---|
| M1 | 1.04 | 3.37 | 8.74 | 5.00 |
| M2 | 1.02 | 3.70 | 9.46 | 5.59 |
| M3 | 1.01 | 3.75 | 9.56 | 5.67 |
| N1 | 1.20 | 2.90 | 7.81 | 4.08 |
| N2 | 1.20 | 3.19 | 8.48 | 4.58 |
| N3 | 1.19 | 3.24 | 8.60 | 4.67 |
| O1 | 1.19 | 2.91 | 7.83 | 4.11 |
| O2 | 1.18 | 3.21 | 8.51 | 4.63 |
| O3 | 1.17 | 3.27 | 8.64 | 4.74 |
| P1 | 0.94 | 3.41 | 8.72 | 5.14 |
| P2 | 0.92 | 3.78 | 9.54 | 5.81 |
| P3 | 0.91 | 3.86 | 9.72 | 5.97 |
| Q1 | 1.12 | 3.09 | 8.19 | 4.47 |
| Q2 | 1.11 | 3.36 | 8.79 | 4.94 |
| Q3 | 1.10 | 3.46 | 9.01 | 5.12 |

Electrophilicity ($\omega$) measures the dye's ability to accept electrons while the electro accepting power ($\omega^+$) and electro donating power ($\omega^-$) quantify the tendencies to donate or accept electron (Delgado-Montiel et al., 2019). The P series dyes also showed the highest $\omega$ and $\omega^+$ of 3.41 to 3.86 eV and 5.14 to 5.97 eV, respectively. This indicates a strong affinity for electrons which is good for electron regeneration by $I^-/I_3^-$. The N and O dyes have the lowest $\omega$ of 2.90 to 3.27 eV and $\omega^+$ values of 4.08 to 4.74 eV. The electron donating ability is largest for P dyes (8.72-9.72 eV) while smallest for N and O dyes (7.81-8.64 eV). In practice, a dye with low $\eta$ combined with high $\omega$ and $\omega^+$ tends to inject electrons easily and be readily reduced by the electrolyte (Delgado-Montiel et al., 2019). Previous study found that dyes with highest $\omega^+$ and lowest $\eta$ produces the greatest short-circuit current and PCE (Badawy et al., 2024). Thus, P and M series dyes should transfer charge rapidly but may risk faster recombination or reduced stability, whereas N and O series dyes may have higher $V_{OC}$ and chemical robustness with low $J_{SC}$.

Importantly, all the dyes have $\eta$ far below than $TiO_2$, implying the dyes can release electrons more easily than the semiconductor. In other words, electron injection from an excited dye into semiconductor conduction band should be thermodynamically favorable (Mujtahid et al., 2022). Similarly, the calculated HOMO energies lie below $I^-/I_3^-$ redox level and the LUMO above the $TiO_2$ band edge, making dye regeneration and injection both spontaneous. In sum, the conceptual descriptors imply that P and M dyes might have very rapid charge transfer and regeneration.

**Photovoltaic Properties**

The light harvesting efficiency (LHE) is directly linked to the oscillator strength ($f$) (Hashmat et al., 2024). Thus, dyes with larges $f$ capture more photons and can drive higher $J_{sc}$. Table 4 shows that, M series dyes have the largest LHE of 0.84 to 0.88 since the $f$ were high for each of the M series dyes, therefore yielding the highest $J_{sc}$ value for M3 (17.53 mA/cm$^2$) (Table 5). In contrast, the Q series dyes have weak LHE and correspondingly low $J_{SC}$ with the lowest value of 8.46 mA/cm$^2$ for Q1. This finding matches the theoretical expectations of showing higher $J_{SC}$ values with higher oscillator strength and LHE.

**Table 4** Light-harvesting efficiency (LHE), ground-state oxidation potential energy ($E_{dye}$), oxidation potential energy of the excited state ($E_{dye}*$), the driving force of regeneration ($\Delta G_{reg}$) and free energy of electron injection ($\Delta G_{inj}$) of studied dyes

| Dye | LHE | $E_{dye}$ | $E_{dye}*$ | $\Delta G_{reg}$ | $\Delta G_{inj}$ |
|---|---|---|---|---|---|
| M1 | 0.84 | 4.77 | 3.03 | -9.57 | 7.03 |
| M2 | 0.87 | 4.89 | 3.25 | -9.69 | 7.25 |
| M3 | 0.88 | 4.90 | 3.27 | -9.70 | 7.27 |
| N1 | 0.75 | 4.93 | 3.18 | -9.73 | 7.18 |
| N2 | 0.80 | 5.10 | 3.49 | -9.90 | 7.49 |
| N3 | 0.81 | 5.12 | 3.51 | -9.92 | 7.51 |
| O1 | 0.70 | 4.90 | 3.24 | -9.70 | 7.24 |
| O2 | 0.78 | 5.06 | 3.50 | -9.86 | 7.50 |
| O3 | 0.79 | 5.07 | 3.48 | -9.87 | 7.48 |
| P1 | 0.69 | 4.52 | 2.73 | -9.32 | 6.73 |
| P2 | 0.77 | 4.65 | 3.00 | -9.45 | 7.00 |
| P3 | 0.78 | 4.66 | 3.02 | -9.46 | 7.02 |
| Q1 | 0.42 | 4.83 | 3.50 | -9.63 | 7.50 |
| Q2 | 0.52 | 4.96 | 3.69 | -9.76 | 7.69 |
| Q3 | 0.53 | 4.99 | 3.72 | -9.79 | 7.72 |

The excited state oxidation potential determines the driving force for electron injection into the semiconductor ($TiO_2$). This value also affects the $\Delta G_{inj}$ parameter which is considered good if $\Delta G_{inj} >> 0.2$ eV (Deogratias et al., 2021). In this study the values are clearly above the minimal threshold for each dye (6.73-7.72 eV) indicating the effectiveness of all the dyes. However, Aftabuzzaman et al (2021) suggested that, excessively large $\Delta G_{inj}$ can reduce $V_{OC}$ and waste energy which can be noticed in case of Q series dyes yielding lower $J_{SC}$ and $V_{OC}$ (8.46 mA/cm$^2$ and 1.40 V, respectively for Q1) despite having large driving force which could be because of thermalization losses (Table 5). By contrast N1 having moderate $\Delta G_{inj}$ (7.18 eV) achieved the highest $V_{OC}$ value of 1.47 V. This balance is analogous to recent findings that optimal dyes provide strong injection driving force without excessive energy loss (Yelkovan et al., 2025). Regeneration free energy ($\Delta G_{reg}$) is uniformly very negative (-9.32 to -9.90 eV in Table 4), indicating dye regeneration by the $I^-/I_3^-$ electrolyte is spontaneous and facile for all dyes. In practice this means the oxidized dye is easily reduced, minimizing recombination at the dye/ electrolyte interface.

**Table 5** Photovoltaic performance of porphyrin-based dyes in DSSC

| Dye | $V_{OC}$ | $J_{SC}$ | PCE (%) |
|---|---|---|---|
| **M1** | 1.30 | 16.90 | 14.28 |
| **M2** | 1.14 | 17.37 | 12.87 |
| **M3** | 1.12 | 17.53 | 12.76 |
| **N1** | 1.47 | 15.03 | 14.37 |
| **N2** | 1.29 | 16.04 | 13.45 |
| **N3** | 1.26 | 16.26 | 13.31 |
| **O1** | 1.47 | 14.05 | 13.42 |
| **O2** | 1.29 | 15.63 | 13.11 |
| **O3** | 1.26 | 15.75 | 12.90 |
| **P1** | 1.36 | 13.70 | 12.11 |
| **P2** | 1.19 | 15.33 | 11.86 |
| **P3** | 1.16 | 15.58 | 11.75 |
| **Q1** | 1.40 | 8.46 | 7.70 |
| **Q2** | 1.25 | 10.39 | 8.44 |
| **Q3** | 1.20 | 10.61 | 8.28 |

Accordingly, the factors together can be taken into accounts to predict the dye with best overall performance. These quantitative indicators not only guide molecular design but also provide parameters (efficiency, voltage, stability markers) that can be fed into techno-economic or system-level models for early-stage technology assessment. Dyes with high LHE and adequate $\Delta G_{inj}$ deliver the highest $J_{SC}$ while the $V_{OC}$ mainly depends on the dye's HOMO level and recombination kinetics. Consequently, the best performing dye in the 15-dye series is the one that achieves an optimal balance among all these properties. N1 attains the highest total PCE of 14.37% through establishing a successful equilibrium between a high open-circuit voltage of 1.47V and sustainable light harvesting efficiency (0.75), along with an adequate thermodynamic driving force for electron injection. Finally, almost all dyes show excellent dye-electrolyte compatibility, HOMO and LUMO energy levels, $\Delta G_{inj}$, $\Delta G_{reg}$ and other optoelectrical parameters while N1 performing outstanding with improved qualities, making this selected dye a perfect candidate for synthesis and practical applications.

**Relevance for Energy System Modelling and Decision Support**

Energy system models are computational tools that are widely utilized for the design and planning of energy systems under technical, economic, and policy constraints. They usually solve optimization problems that minimize total system cost or emissions while meeting energy demand, subject to constraints on technology availability and performance. The input dataset usually includes technology efficiencies, costs (often expressed as Levelized Cost of Electricity, LCOE), and lifetimes, along with demand profiles and resource availability. The model then determines the cost-optimal mix of technologies and infrastructure investments over time. Because these models are widely used to inform policy and R&D strategies, providing accurate and forward-looking technology performance data, such as those generated in this study, can directly improve the quality of transition scenarios and investment decisions.

A unique contribution of this work is its visionary connection between molecular-scale computational chemistry and system-level energy informatics, which has the potential to reduce the gap between material discovery and system-level modelling. The outputs of this study, electronic and optical properties (HOMO–LUMO gap, electromagnetic absorption spectra), charge transfer parameters ($\Delta G_{inj}$, $\Delta G_{reg}$), and photovoltaic performance estimates ($J_{sc}$, $V_{oc}$, $P_{CE}$) provide material-level performance indicators for DSSCs with specific dye candidates. These predicted efficiencies and stability markers can inform broader energy system modelling frameworks in several ways.

First, in bottom-up energy system optimization models such as TIMES, MARKAL, MESSAGE, or OSeMOSYS, technologies are characterized by parameters including efficiency, cost, degradation, and availability (Fodstad et al., 2022). The PCE and stability indicators predicted in this study could be incorporated as input dataset, representing performance assumptions for DSSCs, enabling meaningful comparisons with established PV technologies such as silicon, perovskites, or organic photovoltaics. Second, the relevance of DSSCs to distributed energy and microgrid applications highlights their potential in low-cost, small-scale, or building-integrated contexts. Models of distributed energy resources can use the predicted efficiency values to simulate the contribution of DSSCs to household- or community-scale deployments, particularly in rural or resource-constrained settings.

Third, technology foresight and scenario analyses, including integrated assessment models (IAMs), frequently explore the role of emerging technologies in future energy systems (Pietzcker et al., 2017). Computational predictions like those provided in this study can act as early performance assumptions for DSSCs in forward-looking scenarios and what-if analyses, for example testing the implications of DSSCs reaching a target efficiency at a specific cost by 2035 (e.g., "what if DSSCs reach X% efficiency at Y $/kW by 2035?").

Finally, within decision-support frameworks such as multi-criteria decision analysis (MCDA) or prescriptive analytics for energy R&D investment, the quantitative indicators derived here provide a technical feasibility evidence base (Abo-Zahhad et al., 2024). The parameters derived in this study can help guide decisions about which photovoltaic technologies to prioritize for further research, synthesis, and scale-up. As a proof of concept, Table 6 shows how the predicted PCE values from this study could be translated into simplified levelized cost of electricity (LCOE) estimates under hypothetical assumptions of 75 USD/m² module cost and 10-year lifetime. While purely illustrative, such calculations demonstrate how molecular-level predictions can provide early techno-economic indicators and inform system-level modelling.

**Table 6.** Illustrative back-of-the-envelope LCOE calculation for the top three dye candidates identified in this study

| Dye Candidate | PCE (%) | Module Cost (USD/m²) | Lifetime (years) | Simplified LCOE (USD/kWh)* |
|---|---|---|---|---|
| **N1** | 14.37 | 75 | 10 | 0.11 |
| **M1** | 14.28 | 75 | 10 | 0.11 |
| **N2** | 13.45 | 75 | 10 | 0.12 |

The LCOE values in Table 6 were obtained using a simplified bottom-up calculation that divides the total installed module cost per square meter by the cumulative energy yield over the assumed lifetime of the device. This is a first-order approximation that can be found in several conceptual or early-stage PV assessment papers. It is used for illustrative purposes to provide a conceptual introduction on how to bridge the predictions made in this work, with dataset that can be fed into energy system models.

$$LCOE = \frac{Cost\ per\ m^2}{Annual\ Yield \times Lifetime} \quad (14)$$

The annual energy yield is computed as the product of the predicted PCE values from this work, and a reference solar insolation of 1000 kWh/m²/year, and then multiplied by the device lifetime to give total energy output. The proposed formulation is a first order estimate. Within such formulation, the performance ratio (PR) accounts for temperature effects, mismatch, soiling, inverter losses, etc. and it is often assumed PR ≈ 1.0 for an idealized upper bound.

$$Annual\ Yield = PCE \times Solar\ Insolation \times PR \qquad (15)$$

$$Annual\ Yield = PCE \times 1000 kWh/m^2/year \qquad (16)$$

The proposed LCOE estimates are purely illustrative and rely on idealized assumptions (e.g., PR = 1.0, fixed module cost of 75 USD/m², no BOS cost breakdown, 10-year lifetime). They represent an upper-bound efficiency case and should not be interpreted as real-world performance projections. Although highly simplified, this approach follows the standard definition of LCOE used in techno-economic assessments and provides a first approximation of the cost of electricity generated by DSSCs under idealized conditions. Such LCOE values can be readily incorporated into bottom-up energy system models, where technology costs and efficiencies are key inputs to capacity expansion and investment optimization problems. Importantly, different PCE values lead to different LCOE estimates, which in turn can shift the model's optimal technology mix and thus influence system-level decision making. For example, a DSSC technology with a 14% PCE might compete with other emerging PV technologies in rooftop or building-integrated applications, whereas an 8% PCE option would appear less competitive and therefore receive lower investment share in a capacity expansion model. This highlights how molecular-level predictions can directly affect scenario outcomes and R&D prioritization in energy system planning. Because LCOE is a primary cost metric used in bottom-up energy system models, these results can directly feed into capacity expansion and scenario planning exercises, bridging molecular design with system-level energy transition analyses. Of course, for the PCE to be useful for energy system modelling, they have to be wrapped up in techno-economic assumptions. This is common in prospective studies, where researchers often build scenario analysis with optimistic case (high PCE, low cost, long lifetime); baseline case (realistic manufacturing scale-up); pessimistic case (lower PCE, shorter lifetime). This allows energy system models to explore the impact potential of the new material.

In summary, this work provides an initial step to the increasing need for connecting predictive and prescriptive analytics. Indeed, connecting predictive and prescriptive analytics has been recognized as a key step in enabling data-driven decision-making in energy systems. For instance, predictive tools such as ramping behaviour analyses have been used to extract features from wind power data or partial discharge diagnostics, which then serve as inputs to optimization and planning models for grid operation and asset management (Mishra et al., 2020; Mishra et al., 2024). This integration of forecasting with decision support is essential for building resilient, cost-effective energy infrastructures as highlighted in Mishra et al., (2022). From this point of view, this work contributes to both predictive analytics, by screening and forecasting material/device performance, and prescriptive analytics, by informing decisions on which dyes to synthesize, which PV technologies to scale, and how to integrate DSSCs into future energy portfolios. In doing this, it contributes to the broader domain of energy informatics, where computational methods connect predictive and prescriptive analytics to support sustainable photovoltaic development.

Beyond providing single-point LCOE estimates, the predicted efficiencies from this study can be useful as starting data points for constructing cost-learning trajectories of DSSC technologies. Energy system models often represent emerging technologies with cost reduction pathways following experience curves (learning rates). By combining early-stage performance predictions with assumed manufacturing scale-up scenarios, our pipeline can help define realistic learning curves for DSSCs, enabling scenario analyses that explore their long-term competitiveness under different cost and deployment assumptions.

Overall, this study not only advances the dye design at the molecular scale but it also shows how such predictions can be fed into broader energy system tools, proposing a conceptual pipeline that links molecular-level predictions to techno-economic indicators and system-level modelling inputs, and therefore bridging computational chemistry and energy informatics for sustainable photovoltaic innovation. This study discusses how computational chemistry can serve as a predictive layer in the innovation pipeline, providing data that supports system-level modelling and decision-making for the global energy transition.

**Conclusions**

This study applies DFT/TD-DFT simulations on an HPC-class workstation to investigate fifteen novel porphyrin-based D-π-A organic dyes for dye-sensitized solar cells (DSSCs). Dyes in the M and P series featuring extended π donors have the lowest bandgap and highest electrophilicity which promotes efficient charge transfer. Whereas N series dyes, specially N1 exhibit better optoelectrical properties compared to other dyes achieving a high PCE of 14.37%. These results suggest that pairing a strong donor with a robust anchoring acceptor is a promising structural strategy for maximizing solar cell efficiency. This study proposed techno-economic and system-level modeling frameworks that remained conceptual and were not implemented with a validated dataset. Addressing this gap in future work could involve integrating computationally derived parameters into bottom-up energy system models such as TIMES, MARKAL and OSeMOSYS. Additionally, these parameters could inform distributed energy simulations, microgrid optimization and integrated assessment models to explore long-term energy transition pathways. Furthermore, in-silico screening provides early-stage quantitative insights that fuel predictive analytics for forecasting photovoltaic material and device performance, as well as prescriptive analytics for guiding synthesis, prioritization, and scaling. By integrating computational chemistry with energy informatics and leveraging intra and inter-job parallelization, this approach enables large-scale screening campaigns, accelerates dye discovery and generates rich datasets for machine learning–driven solar innovation.

**Acknowledgements**

The authors wish to express their gratitude to UiT The Arctic University of Norway, specifically the Department of Computer Science, for providing the high-performance computing infrastructure (Dell Precision 7875 Tower workstation) used to execute the DFT and TD-DFT calculations.

**Declaration of Generative AI and AI-Assisted Technologies in the Writing Process**

While preparing this work, the author(s) used Grammarly to paraphrase and edit the language. After using those tools, the author(s) reviewed and revised the content as needed and take(s) full responsibility for the publication's content.

**Declaration of Competing Interest**

The authors declare that they have no known competing financial interests or personal relationships that could have appeared to influence the work reported in this paper.

**Consent to Publish**

All authors agreed with the manuscript's content and gave explicit permission to publish the work.

**Data Availability**

All data used to prepare the manuscript is included. Additional explanations will be available upon request.